\documentclass[prl, superscriptaddress, twocolumn, showpacs,
nofootinbib]{revtex4}

\usepackage[]{graphicx}
\usepackage{amsmath}
\usepackage{amsfonts}
\usepackage{color}
\usepackage{hyperref}
\usepackage{bm}
\usepackage{graphicx}
\usepackage{amsbsy}

\def\ket#1{| #1 \rangle}

\def\bk#1#2{\langle #1 | #2 \rangle}

\newcommand{\aH}{{\widetilde{H}}}
\newcommand{\aZ}{{\widetilde{Z}}}

\pacs{03.67.Pp, 03.67.Lx, 42.65.-k}

\date{\today}

\begin{document}

\title{Stabilizer Quantum Error Correction with Qubus Computation}

\author{Casey R. Myers}\email{crmyers@iqc.ca}
\affiliation{Department of Physics and Astronomy, and Institute for Quantum Computing, University of
Waterloo, ON, N2L 3G1, Canada}
\author{Marcus Silva}\email{msilva@iqc.ca}
\affiliation{Department of Physics and Astronomy, and Institute for Quantum Computing, University of
Waterloo, ON, N2L 3G1, Canada}
\author{Kae Nemoto}
\affiliation{ National Institute of Informatics, Hitotsubashi,
Chiyoda-ku, Tokyo 101-8430, Japan}
\author{William J. Munro}
\affiliation{Hewlett-Packard Laboratories, Filton Road, Stoke
Gifford, Bristol, BS34 8QZ, UK}
\affiliation{ National Institute of Informatics, Hitotsubashi,
Chiyoda-ku, Tokyo 101-8430, Japan}

\begin{abstract}

In this paper we investigate stabilizer quantum error correction codes
using controlled phase rotations of strong coherent probe states. We
explicitly describe two methods to measure the Pauli operators which
generate the stabilizer group of a quantum code. First, we show how to
measure a Pauli operator acting on physical qubits using a single
coherent state with large average photon number, displacement
operations, and photon detection. Second, we show how to measure the
stabilizer operators fault-tolerantly by the deterministic preparation of
coherent cat states along with one-bit teleportations between a
qubit-like encoding of coherent states and physical qubits.

\end{abstract}

\maketitle

The question of which physical system is best suited for quantum information
processing is still open, each implementation proposal having
strengths and weaknesses. In some systems (such as optics) 
it is difficult to make qubits interact, so that the two-qubit gates
needed for universal computation are difficult to implement.
One scheme, proposed by Gottesman {et. al.}~\cite{Gottesman99},
circumvents the need to make qubits interact directly by using a modified
teleportation protocol. A generalization of this leads to the cluster state
proposal of Raussendorf {\em et. al.}~\cite{Raussendorf01}, where a
large entangled state is prepared offline, and computation is
performed by a sequence of single qubit measurements 
which depend on the outcomes of previous measurements. 
A different scheme that bypasses the need for qubits to interact directly
was proposed by Nemoto {\em et. al.}~\cite{Nemoto04,Nemoto05}. This scheme shows
that, by inducing a phase on a large coherent state bus mode which depends 
on the logical state of the physical qubits, one can implement a near deterministic 
CNOT gate between the qubits. Coherent states are particularly useful 
because of the ease with which they may be produced, e.g. with lasers
or Bose-Einstein condensates. Further developments have
shown more direct methods to perform two-qubit gates with bus
modes, termed {\em qubus computation}~\cite{Spiller06}.  

If qubus computation is to be seriously considered for
physical implementation, a full analysis of the propagation of errors
should be undertaken. The starting point for these considerations 
is whether we can perform quantum error correction (QEC) on qubits efficiently. 
In particular, can one measure the syndromes for a given stabilizer code 
directly with controlled rotations (CRs) and strong coherent probe beams? Recent
work by Yamaguchi \textit{et. al.}~\cite{Yamaguchi05} demonstrates how to
measure the syndromes for {\em some} stabilizer codes using these tools. 
They show that the stabilizers for the three
bit-flip code can be measured directly with CRs and a single strong
coherent bus mode. The stabilizers for Shor's 9-qubit
code~\cite{Shor96} can also be measured, showing that it is possible
to correct for any error on a single qubit in an encoded block.

The purpose of this paper is to generalise the results of Yamaguchi {\it et. al}
and demonstrate how CRs can be used to implement quantum error correction with 
{\em any} possible stabilizer code. We will describe two
schemes to measure the syndromes of an arbitrary weight $n$ Pauli
operator, using the stabilizer operators of the seven qubit code as a
concrete example for each one of these schemes.  The first scheme uses
a single strong coherent probe beam, a quadratic number of CRs, a
linear number of coherent displacements, and a photon number
measurement.  This scheme can be modified to use homodyne measurement
at the cost of a slightly larger number of CRs and coherent
displacements. The second scheme we describe is a fault-tolerant
approach to the measurement of the Pauli operators, which requires a
linear number of strong coherent pulses, CRs and detectors. Although
we focus on the 7 qubit code -- which  has stabilizer generators with
weight 4 -- for each of these schemes we describe how to generalise to
Pauli operators of weight $n$.

\begin{figure}[ht]
\includegraphics[width=7cm]{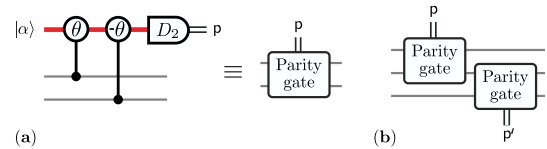}
\caption{\footnotesize {\bf (a)} Circuit to measure the parity of
two 
qubits. The CRs are $\pm\theta$.  {\bf (b)} Two parity gates combined to measure the Pauli operators $ZZI$ and $IZZ$.
\label{YamaguchiGHZ}}
\end{figure}

{\em Background ---} 
In~\cite{Yamaguchi05} it was shown that the stabilizers for the
3-qubit bit-flip code ($\ket{0}$$\to$$\ket{000}$, $\ket{1}$$\to$$\ket{111}$)
could be measured with the parity gate depicted in Fig.~\ref{YamaguchiGHZ}a.

It can be seen that this circuit is a parity gate when we consider its
effect on the input state
$\ket{\psi_{\text{in}}}=
\bigl(a_{0}\ket{00}+a_{1}\ket{01}+
a_{2}\ket{10}+a_{3}\ket{11}\bigr)\ket{\alpha}$,
where $\ket{\alpha}$ is a coherent bus mode. 
The effect of the CRs is to  apply a phase to the coherent beam
if our data qubit is $\ket{1}$ and leave it alone otherwise:
$\bigl(a\ket{0}+b\ket{1}\bigr)\ket{\alpha}\to
a\ket{0}\ket{\alpha}+b\ket{1}\ket{\alpha e^{i\theta}}$. Before
the detector $D_{2}$ in Fig.~\ref{YamaguchiGHZ}a, the state
$\ket{\psi_{\text{in}}}$ is
$\bigl(a_{0}\ket{00}+a_{3}\ket{11}\bigr)\ket{\alpha}+
a_{1}\ket{01}\ket{\alpha e^{-i\theta}}+a_{2}\ket{10}\ket{\alpha e^{i\theta}}$. When
$D_{2}$ is a homodyne detection along the $x$-quadrature we are able
to distinguish $a_{0}\ket{00}+a_{3}\ket{11}$ from
$a_{1}\ket{01}+a_{2}\ket{10}$, since a homodyne measurement
of $\ket{\alpha}$ along the $x$-quadrature is equivalent to the
projection $\bk{x}{\alpha}$.  
That is, $\ket{\alpha e^{\pm i\theta}}$ are indistinguishable when we homodyne detect along
the $x$-quadrature. This is the basis of the CNOT shown in~\cite{Nemoto04}.

With two parity gates we can measure the Pauli operators $ZZI$ and
$IZZ$. That is, one parity gate is applied to qubits 1 and 2 to
measure $ZZI$ while the second parity gate is applied to qubits 2 and
3 to measure $IZZ$, as shown in Fig.~\ref{YamaguchiGHZ}b. The state
before the application of the parity gates is
$\ket{\overline{\psi}_{\text{in}}}=
\bigl(c_{0}\ket{000}+c_{1}\ket{111}\bigr)\ket{\alpha}\ket{\alpha}$. There
are four cases to consider:  no error,
$\ket{\overline{\psi}_{\text{in}}}$; an error on qubit 1,
$XII\ket{\overline{\psi}_{\text{in}}}$; an error on qubit 2, $I
XI\ket{\overline{\psi}_{\text{in}}}$; an error on qubit 3, $II
X\ket{\overline{\psi}_{\text{in}}}$. We can see what the effect of a
bit flip error on each of the modes is by considering the state
$\ket{abc}\ket{\alpha}\ket{\alpha}$, where $a,b,c\in
\{0,1\}$. Directly before homodyne detection in
Fig.~\ref{YamaguchiGHZ}b $\ket{abc}\ket{\alpha}\ket{\alpha}$ becomes
$\ket{abc}\ket{\alpha e^{i(a-b)\theta}}\ket{\alpha e^{i(c-b)\theta}}$. When
we measure the probe states to be $\ket{\alpha e^{\pm im\theta}}\ket{\alpha e^{\pm in\theta}}$, 
where $m,n\in\{0,\pm1\}$, we know whether there was no error ($m,n=0$) or a
one bit flip error, the location of the bit flip also being identified
by the values of $m$ and $n$. Similar methods can be applied to
measure the stabilizer operators for Shor's 9-qubit code. The natural
question that arises is: can we use techniques similar to those above
to measure the syndromes for an arbitrary stabilizer code?

{\em Larger codes ---}
As a concrete example, consider the $[[7,1,3]]$ stabilizer 
code~\cite{Steane}. This code can correct a single arbitrary 
quantum error in any of the 7 qubits, and it has been used extensively in studies of 
fault-tolerance in quantum computers due to the fact that it allows for simple 
constructions of fault-tolerant encoded gates~\cite{Got98}. In order to detect
which error has corrupted the data, one must measure six multiqubit
Pauli operators which, up to qubit permutations and local unitaries, 
are equivalent to the Pauli operator $ZZZZ$, or the measurement of
{\em only} the parity of 4 qubits. For an arbitrary stabilizer code, 
various multiqubit Pauli operator must be measured, each of which is
always equivalent to a measurement of only the parity of a subset
of qubits, thus it is sufficient to consider only
multiqubit parity measurements in order to perform quantum error
correction with stabilizer codes.

{\em Single Coherent State Pulse ---} 
In order to measure $ZZZZ$ with CRs, 
we can start with the encoded state $\bigl(c_{0}\ket{0_{L}}+c_{1}\ket{1_{L}}\bigr)\ket{\alpha}$, and 
design a circuit that gives us $\ket{\alpha_{1}}$ when there was no error (even
parity) and $\ket{\alpha_{2}}$ when there was an error (odd parity),
where $\alpha_{1}\neq \alpha_{2}$.  

\begin{figure}[ht]
\includegraphics[width=7cm]{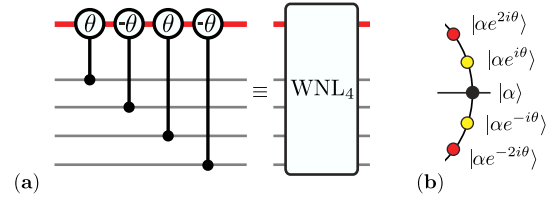}
\caption{\footnotesize {\bf (a)} First attempt at using CRs to measure the parity of four qubits with a single strong
coherent probe. {\bf (b)} Position of phase space peaks corresponding
to the state of the coherent probe beam. Yellow circles correspond to odd states while red and black circles correspond to even states.
\label{4qubitNoDisp4qubitNoDispPhase}}
\end{figure}

Ideally we would want to do this with just one coherent probe beam, four
CRs and a single homodyne detection, following a
direct analogy with the circuit depicted in Fig.~\ref{YamaguchiGHZ}a. 
However this is
not possible. The best we can do is have some even states go to
$\ket{\alpha}$ and the rest go to $\ket{\alpha e^{\pm2i\theta}}$ while
the odd states go to $\ket{\alpha e^{\pm i\theta}}$. The circuit that
performs this is shown in Fig.~\ref{4qubitNoDisp4qubitNoDispPhase}a.

Notice that in phase space we would have five points -- three for the even
states ($\ket{\alpha}, \ket{\alpha e^{\pm2i\theta}}$) and two for the
odd states ($\ket{\alpha e^{\pm i\theta}}$) -- as can be seen in
Fig.~\ref{4qubitNoDisp4qubitNoDispPhase}b. If we were to homodyne detect the probe beam at this stage, we would
partially decode our encoded state $c_{0}\ket{0_{L}}+c_{1}\ket{1_{L}}$
since we can distinguish the state $\ket{\alpha}$ from
$\ket{\alpha e^{\pm2i\theta}}$. The problem now becomes determining
what operations must be done before we homodyne detect so that 
we only distinguish 
between states of different parity in the first four qubits, 
and nothing more. It turns out that either homodyne or photon number detection can be used,
depending on the operations applied before the measurement.

\begin{figure}[ht]
\includegraphics[width=5cm]{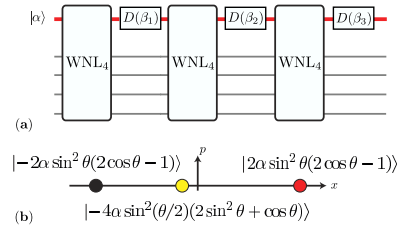}
\caption{\footnotesize {\bf (a)} CRs used to measure the parity of four qubits with a photon number detection. {\bf (b)}  Position of phase space peaks corresponding to the state of the coherent probe beam.
\label{4qubit2DispPhase}}
\end{figure}

{\em Photon number measurement ---}
If we incorporate displacements along with Fig.~\ref{4qubitNoDisp4qubitNoDispPhase}a
we can take the five points in phase space to just
three. Displacements of a state can be easily implemented by mixing
the state with a large coherent state on a weak beam splitter, the
size of the coherent states amplitude and beam splitter reflectivity
deciding the displacement~\cite{Displ}. If we have three displacements and three applications of Fig.~\ref{4qubitNoDisp4qubitNoDispPhase}a, as in Fig.~\ref{4qubit2DispPhase}a, we find
that $\ket{\text{odd}}\to\ket{-4\alpha\sin^{2}(\theta/2)(2 \sin^{2}(\theta)+\cos(\theta))}$ and
$\ket{\text{even}}\to\ket{\pm 2\alpha\sin^{2}(\theta)(2\cos(\theta)-1)}$, as
depicted in Fig.~\ref{4qubit2DispPhase}b. 
The displacements that accomplish this are
$D(\beta_{1})=D( -4\alpha \cos^{2}(\theta/2)\bigl(2\cos(\theta)-1\bigr))$, $D(\beta_{2})=D(\alpha\bigl(1+2\cos(\theta)+2\cos(3\theta)\bigr))$ and $D(\beta_{3})=D(\alpha(\cos(2\theta)-\cos(3\theta)-\cos(\theta)-1)$. 

Notice that the red and black circle in the  Fig.~\ref{4qubit2DispPhase}b are equidistant from the p-axis. We can thus perform a photon
number measurement on the probe beam to determine whether we had an odd or even
state. 
In order for a photon number measurement to distinguish the odd from even states we require $\alpha\theta^{2}\gg 1/\sqrt{3}$. 

We can use this method to measure the parity for a state of any
size. If we have $n$ qubits then we can have at best $n+1$ points in
phase space, using the $\theta, -\theta, \theta, -\theta$ pattern for
the CRs shown in Fig.~\ref{4qubitNoDisp4qubitNoDispPhase}a. Using
displacements and a photon number detector we are able to measure
the parity.  In general, if $n$ is even we need $n-1$ displacements
and $n^{2}-n$ CRs with a photon number measurement. When $n$ is odd, after the application of the circuit analogous to Fig.~\ref{4qubitNoDisp4qubitNoDispPhase}a of size $n$, we will have the point $\ket{\alpha e^{i(2n-1)\theta}}$ in phase space without the point $\ket{\alpha e^{-i(2n-1)\theta}}$. So we need an extra displacement to move the non-symmetric point. If $n$ is odd we need $n$ displacements and $n^{2}$ CRs with a photon number measurement. 

For this method to work we need the use of a number discriminating
photo-detector. In practice it is well known that homodyne detection is much more precise
than number discriminating photo-detectors. For this reason, we describe how to measure a Pauli operator 
using homodyne detection.

{\em Homodyne detection ---} Consider the $ZZZZ$ case again. After applying 
Fig.~\ref{4qubitNoDisp4qubitNoDispPhase}a we have five points in phase space. Ideally we want $\ket{\alpha}$ and $\ket{\alpha e^{\pm2i\theta}}$ to become one point in phase space, say  $R_{1}+iR_{2}$, and $\ket{\alpha e^{\pm i\theta}}$ to become one point, say  $R_{3}+iR_{4}$. If this was possible then homodyne detection could be used. This can be done with five displacements and six applications of Fig.~\ref{4qubitNoDisp4qubitNoDispPhase}a, requiring 10 simultaneous equations to be solved. Without loss of generality we set $R_{2}=R_{4}=R_{1}=0$. The equations to be solved are $e^{4i\beta\theta}A+e^{3i\beta\theta}B+e^{2i\beta\theta}C+e^{i\beta\theta}D+E=e^{-i\beta\theta}\alpha_{\beta}-e^{5i\beta\theta}$, where $\alpha_{0,\pm2}=0$ and $\alpha_{\pm 1}=R_{3}$.

After solving these equations we find that $A$, $B$, $C$, $D$ and $E$ scale as
$-R_{3}/\theta^{4}$. We are free to choose the distance between the origin and $R_{3}$ to be arbitrarily large, at the expense of using arbitrarily large displacements.
We can also use the above method to distinguish the parity of any
given state of $n$ qubits. If we have $n$ qubits we have $n+1$ points
in phase space, using a circuit similar to Fig.~\ref{4qubitNoDisp4qubitNoDispPhase}a. 
In order to distinguish the parity we need $n+1$ displacements and $n(n+2)$ CRs.

{\em Fault-Tolerance ---} 
These two methods to measure weight $n$ Pauli
operators cannot be used for fault-tolerant quantum computation. 
If there is an error on the
coherent probe mode during one of the CRs, 
say photon loss, it would be transferred to a phase
error in each of the physical qubits it interacts with afterwards --
that is, a single fault can cause a number of errors which 
is greater than the number of errors the code can correct.
For this reason we now look at
measuring the syndromes of stabilizers fault-tolerantly.

\begin{figure}[ht]
\includegraphics[width=8.3cm]{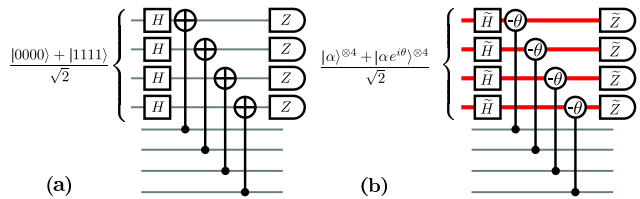}
\caption{\footnotesize {\bf(a)} Circuit for the measurement
of the parity of four qubits~\cite{Shor96}. {\bf(b)} Same circuit
modified to use coherent states and CRs.
\label{ShorCirc}} 
\end{figure}

Shor~\cite{Shor96}
first described how to fault-tolerantly measure the generators of the
stabilizer group of a quantum error correcting code using ancilla GHZ states
$\bigl(\ket{0}^{\otimes n}+\ket{1}^{\otimes n}\bigr)/\sqrt{2}$, CNOT's
and Hadamards. For example, in order to measure the Pauli operator
$ZZZZ$ (which is equivalent to measuring 
the parity of 4 qubits and nothing else), we would use the circuit 
shown in Fig.~\ref{ShorCirc}a.

To fault-tolerantly measure the stabilizer group generators of a QEC with CRs we make three modifications to Fig.~\ref{ShorCirc}a. First,
instead of using $\ket{0}$ and $\ket{1}$ for the ancilla, 
we use the coherent states $\ket{\alpha}$ and
$\ket{\alpha e^{i\theta}}$, respectively. In that case, the ancilla GHZ state becomes
$\bigl(\ket{\alpha}^{\otimes n}+\ket{\alpha e^{i\theta}}^{\otimes n}\bigr)/\sqrt{2}$. 
Second, we replace the CNOT's 
with CRs, which will cause a phase shift if the
physical state is $\ket{1}$ and do nothing otherwise, i.e. $\ket{1}\ket{\alpha}\to\ket{1}\ket{\alpha e^{-i\theta}}$ and $\ket{1}\ket{\alpha e^{i\theta}}\to\ket{1}\ket{\alpha}$. We also need to replace the Hadamards with some quantum operation $\aH$ which will 
perform the mapping
$\aH\ket{\alpha}\approx
\bigl(\ket{\alpha}+\ket{\alpha e^{i\theta}}\bigr)/\sqrt{2}$
and
$\aH\ket{\alpha e^{i\theta}}\approx
\bigl(\ket{\alpha}-\ket{\alpha e^{i\theta}}\bigr)/\sqrt{2}$. 
Third, we replace the qubit measurements with some sort of optical 
measurement that distinguishes between $\ket{\alpha}$ and $\ket{\alpha e^{\pm i\theta}}$
but not between $\ket{\alpha e^{i\theta}}$ and $\ket{\alpha e^{-i\theta}}$ -- this
is what we call $\aZ$ measurement.
This new circuit is depicted in Fig.~\ref{ShorCirc}b.

\begin{figure}[ht]
\includegraphics[width=6cm]{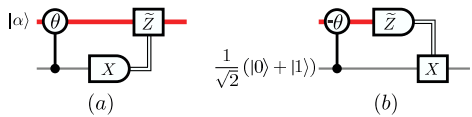}
\caption{\footnotesize Approximate one bit teleportation 
protocols~\cite{Zhou00} using CRs. The measurement
dependent Pauli corrections need not be performed, as discussed
in the body of the paper.
\label{one-teleWNL}}
\end{figure}

The $\aZ$ measurements can be performed directly by homodyne detection, or
by displacements followed by photon counting detectors -- in both cases,
using techniques outlined earlier in this paper.
What remains to be specified is the preparation of the
coherent cat state 
$(\ket{\alpha}^{\otimes n}+\ket{\alpha e^{i\theta}}^{\otimes n})/\sqrt{2}$ and the implementation of the $\aH$ operation. 
One solution for the cat state preparation is to use one 
bit teleportations~\cite{Zhou00}
which translate states from the $\ket{0}/\ket{1}$ basis to the $\ket{\alpha}/\ket{\alpha e^{\pm i\theta}}$ basis. Preparation of the cat state is done by using the one bit teleportation
in Fig.~\ref{one-teleWNL}a to prepare $\ket{\sqrt{n}\alpha}+\ket{\sqrt{n}\alpha e^{i\theta}}$ from the state $\bigl(\ket{0}+\ket{1}\bigr)/\sqrt{2}$ and the coherent state $\ket{\sqrt{n}\alpha}$, and then sending
this state into an $n$-port symmetric beam-splitter~\cite{Gilchrist04,Ralph03}.
In principle, we are required to correct the state before the beam-splitter
by applying the transformation $\aZ$ such that
$\aZ\ket{\alpha}\approx\ket{\alpha}$ while 
$\aZ\ket{\alpha e^{i\theta}}\approx-\ket{\alpha e^{i\theta}}$. However,
we can avoid explicitly applying this transformation by keeping track 
of this necessary correction -- what is called the {\em Pauli 
frame}~\cite{PauliFrame}-- and compensating for 
it in subsequent measurements.
Similarly, to perform the $\aH$ (the approximate Hadamard on
coherent state logic) we
first use Fig.~\ref{one-teleWNL}b to teleport the quantum state 
from the bus to a qubit, 
then perform the Hadamard transformation
and finally teleport back to the coherent state 
logic using the circuit shown in
Fig.~\ref{one-teleWNL}a.

These teleportations,
when performed back-to-back to teleport a qubit state to another qubit,
can also be used as {\em leakage reduction units} to reduce leakage faults
to regular faults~\cite{AT07}.

The resources required to measure a weight $n$ Pauli operator are $3n+1$
CRs, $n+1$ ancillary qubits, $2n$ $\aZ$ measurements
and $n+1$ qubit measurements.

{\em Noisy ancillas --- }  
If the probability of error at each gate is bounded by $\epsilon$,
transversal operations and encoding can ensure that the probability of an 
uncorrectable error is $O(\epsilon^2)$ instead of $O(\epsilon)$.
An error during cat state preparation may lead to correlated $X$-like
errors in the cat state with probability $O(\epsilon)$, which can lead 
to uncorrectable errors in the encoded data during the measurement
of the Pauli operator, thus defeating the purpose of encoding the data 
for fault-tolerant quantum computation. 
In order to avoid this, one can verify the integrity of the cat state
via non-destructive state measurement~\cite{Pre97,AGP}. When using
CRs and coherent beam probes, this translates
to preparing an extra copy of the cat state, which remains in
coherent state logic, interacting with the qubit GHZ state
transversally with controlled $-\theta$ rotations, and $\aZ$ 
measuring each mode of the ancillary cat state. 
By performing classical error correction on the measurement outcomes, 
one can deduce the locations of
$X$-like errors in either the GHZ state or the ancillary cat state. 
If the data is encoded in a code that can correct a single error, 
repeating this procedure with another ancillary cat state allows 
for the inference of which locations in the qubit GHZ state
have $X$ errors with high enough probability to ensure 
uncorrectable errors are only introduced into the data with 
probability $O(\epsilon^2)$~\cite{Pre97}, so that Pauli measurements
with a verified ancilla can be used for fault-tolerant quantum computation.
Overall, the overhead for each attempt of measuring a weight $n$ 
Pauli operator consists of $2(n+1)$ CRs, $2$ ancillary 
qubit preparations and measurements, and $2n$ $\aZ$ measurements. 

$Z$-like errors (including dephasing of coherent superpositions,
one of the consequences of photon loss in the CRs) 
do not lead to errors in the encoded data, just
errors in the outcome of the Pauli operator measurement. If
error correction is to be performed, the Pauli operator measurement
must be repeated $3$ times, and a majority 
vote of the outcomes is taken, in order to ensure that the
measurement outcome is reliable~\cite{Pre97}.

Some of the systematic errors in the probe beams, such as phase rotation
or attenuation (also consequences of photon loss in the CRs), 
can be partially compensated for by additional linear-optics
elements and by adjusting the $\aZ$ measurements individually
to minimize additional $X$ errors. Moreover, errors in the transversal
operations during the preparation of the cat state are independent, and 
thus do not need special consideration during this verification stage --
they do contribute to $\epsilon$, however, and are thus crucial for 
fault-tolerance threshold calculations.

{\em Discussion ---}  We have shown two schemes to measure the
syndromes of an arbitrary weight $n$ Pauli operator. The first scheme
uses a single strong coherent probe beam, a quadratic number of CRs, a
linear number of coherent displacements, and a photon number or
homodyne measurement -- however, this scheme is not fault-tolerant. 
The second scheme we described is fault-tolerant, and the 
amount of resources scales linearly with the weight of the Pauli 
operator. This demonstrates how it is in principle possible to
perform general fault-tolerant quantum computation in the qubus 
architecture. It is worth noting that we could have easily used controlled displacements in the place of CRs in the methods presented here.

{\em Acknowledgments ---} We would like to thank P. Aliferis and R. Van Meter for valuable discussions.  We are supported in part by  NSERC, ARO, CIAR, MITACS, the Lazaridis Fellowship, MEXT in Japan and the EU project QAP.


\begin{thebibliography}{99}

\bibitem{Gottesman99} D. Gottesman and I.L. Chuang, Nature {\bf 402}, 390 (1999)

\bibitem{Raussendorf01} R. Raussendorf and H.J. Briegel, Phys. Rev. Lett. {\bf 86}, 5188 (2001)

\bibitem{Nemoto04} K. Nemoto and W.J. Munro, Phys. Rev. Lett. {\bf 93}, 250502 (2004)

\bibitem{Nemoto05} K. Nemoto and W.J. Munro, Phys. Lett. A {\bf 344}, 104 (2005)

\bibitem{Spiller06} T. P. Spiller {\em el. al. }, New J. Phys. {\bf 8}, 30 (2006)

\bibitem{Yamaguchi05} F. Yamaguchi {\em et. al}, Phys. Rev. A {\bf 73}, 060302 (2006).


\bibitem{Steane} A. Steane, Proc. Roy. Soc. London, Ser. A {\bf 452}, 2551 (1996).

\bibitem{Got98} D. Gottesman, Phys. Rev. A {\bf 57}, 127--137 (1998).

\bibitem{Displ} D.F. Walls and G.J. Milburn, {\em Quantum Optics}, (Springer-Verlag) 1994.


\bibitem{Shor96} P.W. Shor, in Proceedings of the 37th Symposium on Foundations of Computer Science (IEEE, Los Alamitos, 1996), p. 56, quant-ph/9605011. 

\bibitem{Zhou00} X. Zhou {\em et. al}, Phys. Rev. A {\bf 62}, 052316 (2000).

\bibitem{Gilchrist04} A. Gilchrist {\em et. al}, J. Opt. B:  Quantum Semiclass. Opt. {\bf 6}, S828 (2004).

\bibitem{Ralph03} T.C. Ralph {\em et. al}, Phys. Rev. A {\bf 68}, 042319 (2003).

\bibitem{PauliFrame} A. M. Steane, Phys. Rev. A {\bf 68}, 042322 (2003). 
E. Knill, Phys. Rev. A {\bf 71}, 042322 (2005).

\bibitem{AT07} P. Aliferis and B. M. Terhal, Quant. Inf. Comp. {\bf 7}, 139--156 (2007).

\bibitem{Pre97} J. Preskill, Arxiv preprint quant-ph/9712048.

\bibitem{AGP} P. Aliferis {\em et. al}, Quant. Inf. Comp. {\bf 6}(2), 97--165 (2006).

\end{thebibliography}
\end{document}